\documentclass[12pt,a4paper]{article}
\usepackage[utf8]{inputenc}
\usepackage[plainpages=false,pdfpagelabels=true]{hyperref}
\usepackage{amssymb,amsthm}
\usepackage[margin=1.5 in]{geometry}
\usepackage{graphicx}
\usepackage{a4wide}  
\usepackage{amsfonts}
\usepackage{amsmath}
\usepackage{bbm}
\usepackage{booktabs} 
\usepackage{multicol}
\usepackage{ifpdf}

\ifpdf

\hypersetup{%
  pdftitle    = {A Riemannian geometric approach for timelike and null spacetime geodesics}
  pdfkeywords = {Jacobi metric, geodesics, Kerr spacetime, constants of motion},
  pdfauthor   = {Oscar},
  plainpages  = true,
  colorlinks  = true,
  citecolor   = blue,
  urlcolor    = blue,
  linkcolor   = black
}

\newcommand{\arxiv}[1]{{\tt
\href{http://www.arXiv.org/abs/#1}{#1}}}
\newcommand{\doi}[1]{{\tt DOI:\href{http://dx.doi.org/#1}{#1}}}

\makeatletter
\@addtoreset{equation}{section}
\makeatother

\begin{document}

\begin{center}

{\Large {\bf A Riemannian geometric approach for timelike and null spacetime geodesics}}

\vspace{1.5cm}

\renewcommand{\thefootnote}{\alph{footnote}}
{\sl\large  Marcos A. Argañaraz\footnote{E-mail: marcos.arganaraz [at] unc.edu.ar} and Oscar Lasso Andino}\footnote{E-mail: {oscar.lasso [at] udla.edu.ec}}

\setcounter{footnote}{0}
\renewcommand{\thefootnote}{\arabic{footnote}}

\vspace{1.5cm}

{\it $^a$Facultad de Matemática Astronomía, Física y Computación (FaMAF), Universidad Nacional de Córdoba, Ciudad Universitaria, (5000) Córdoba, Argentina.}
\ \vspace{0.3cm}

{\it $^b$ Escuela de Ciencias Físicas y Matemáticas, Universidad de Las Américas,\\
Redondel del Ciclista, Antigua Vía a Nayón, C.P.170124, Quito, Ecuador}\\ \vspace{0.3cm}

\vspace{1.8cm}


{\bf Abstract}

\end{center}

\begin{quotation}
The geodesic motion in a Lorentzian spacetime can be described by trajectories in a $3-$dimensional Riemannian metric. In this article we present a generalized Jacobi metric obtained from projecting a Lorentzian metric over the directions of its Killing vectors. The resulting Riemannian metric inherits the geodesics for asymptotically flat spacetimes including the stationary and axisymmetric ones. The method allows us to find Riemannian metrics in three and two dimensions plus the radial geodesic equation when we project over three different directions. The $3-$dimensional Riemannian metric reduces to the Jacobi metric when static, spherically symmetric and asymptotically flat spacetimes are considered. However, it can be calculated for a larger variety of metrics in any number of dimensions. We show that the geodesics of the original spacetime metrics are inherited by  the projected Riemannian metric. We calculate the Gaussian and geodesic curvatures of the resulting $2-$dimensional metric, we study its near horizon and asymptotic limit. We also show that this technique can be  used for studying the violation of the strong cosmic censorship conjecture in the context of general relativity.
\end{quotation}

\newpage
\pagestyle{plain}


\newpage


\section{Introduction}

The study of geodesics in a given spacetime can be a very hard problem. The trajectories followed by massive particles can be quite complicated and developing new tools for attacking these kind of problems  is an active field of research. In this article we shed light on the problem of describing geodesic motion in Lorentzian spacetimes using Riemannian Geometry.  Globally, the geometry of a spacetime is Lorentzian and the geodesics behave quite different compared with those defined in a Riemannian manifold. For example, in Lorentzian spacetimes there is not an equivalence between different types of geodesic completeness. It would be really helpful to be able to use some results of the Riemannian geometry for studying the geodesics of a spacetime.\\
In a pioneering  work \cite{Pin:1975} it was shown how to geometrize the classical Newtonian dynamics, thus, the dynamics can be translated to a Riemannian manifold and solved by using the tools of the Riemannian geometry. The author uses this new Riemannian metric for studying the classification of trajectories of the N-body problem. We can think about similar problems but concerning Lorentzian spacetimes. We would like to know if by a similar procedure we can study the geodesics defined on a Lorentzian spacetime but using the techniques developed for Riemannian manifolds. A method used for obtaining a Riemannian metric has been developed for static  spacetimes \cite{Gibbons:2015qja}, see also \cite{Szydlowski:1996}. The Riemannian metric obtained by projecting over surfaces of constant energy is called the Jacobi metric. One of the main characteristics of this Jacobi metric is that it inherits the geodesic properties of the original spacetime, therefore we can study the particle dynamics only by knowing the Jacobi metric, a Riemannian metric.\\ 
The extension to time dependent and stationary metrics were studied in \cite{Chanda:2016aph,Chanda:2019guf,Duenas-Vidal:2022kcx}. The formalism was also applied to charged  black holes \cite{Das:2016opi} and  wormholes \cite{Arganaraz:2019fup}. The calculation of the Jacobi metric  worked well when dealing with static, spherically symmetric, asymptotically flat spacetimes and also with time dependent asymptotically flat metrics. However, when applied to stationary spacetimes, these  methods do not give a unique Jacobi metric that inherits the dynamics \cite{Chanda:2019guf}.\\
In this article we introduce a \textit{generalized Jacobi metric} that gives the expected results for stationary spacetimes, having the advantage that this metric is Riemannian\footnote{The Jacobi metric for stationary spacetimes proposed in the literature is of the Randers-Finsler type \cite{Chanda:2019guf}.}. This new metric reduces to the known Jacobi metric when restricted to static or time dependent Lorentzian metrics. The method is general enough for allowing us to calculate a generalized Jacobi metric by projecting over any Killing vector of the metric. When this Killing  vector is $\partial_t$ we have a projection over a surface of constant energy, and then, we obtain a  Jacobi metric that coincides with the one defined in \cite{Gibbons:2015qja}. On the other hand , when the projection is made over two killing vectors a reduced 2-dimensional Jacobi metric that inherits the information regarding the geodesics of spacetime is obtained. We are going to use this result for discussing the stability of null geodesics.\\
We explicitly show that the geodesics of a general stationary spacetime metric are inherited by the generalized Jacobi metric. As a paradigmatic example we take the Kerr metric. We calculate its generalized Jacobi metric and from it we show how to recover the spacelike, null and timelike geodesics.\\
In section \ref{sec:1} we present the generalized Jacobi metric, we deduce it from Hamilton's equation. We compare it with previous results and show that the generalized Jacobi metric reduces to the known cases when calculated for static or dynamic asymptotically flat spacetimes. In section \ref{sec:2} we calculate the generalized Jacobi metric for static, spherically symmetric spacetimes by projecting over two Killing vectors. We also discuss the circular geodesics stability. In section \ref{sec:3} we calculate the generalized Jacobi metric for stationary spacetimes and we discuss its circular geodesics stability. In section \ref{sec:4} we show that the generalized Jacobi metric for stationary spacetime metrics inherits the geodesics of the Lorentzian spacetime metric. We show that the equations of motion calculated with the generalized Jacobi metric are reparametrizations of the equations of motion of the Lorentzian spacetime metric. Using the generalized Jacobi metric we recover the geodesic equation for the Kerr spacetime in section \ref{sec:5}. In section \ref{sec:6} we calculate the geodesic and Gaussian curvature of the $2-$dimensional Jacobi metric and we present the reduced generalized Jacobi metric for different black hole metrics. Finally, in section \ref{discussion} we present the discussion section.

\section{\label{sec:1} The generalized Jacobi metric}

Let us consider a static Lorentzian metric

\begin{equation}\label{metric1}
ds^2=-V^2dt^2+g_{ij}dx^idx^j,
\end{equation}

the corresponding Jacobi metric $J_{ij}$ is defined by 

\begin{equation}
J_{ij}dx^{i}dx^{j}=(E^2-m^2V^2)V^{-2}g_{ij}dx^{i}dx^{j},
\end{equation}

this Jacobi metric is a Riemannian metric and it encodes the geodesic information of the original metric restricted to a surface of constant energy. In \cite{Gibbons:2015qja} the Jacobi metric for  the Schwarzschild spacetime was presented. Due to the fact that for time dependent metrics it is not possible to define a surface of constant energy, the Eisenhart-Duval lift procedure has to be used\cite{Duenas-Vidal:2022kcx}. However, the Jacobi metric provided for stationary spacetimes does not inherit the goedesics of the spacetime metric. In \cite{Chanda:2019guf} the authors provide a different Jacobi metric for stationary spacetimes, but neither does this metric inherit all the geodesic properties.\\
Here we present a new metric that we have called \textit{generalized Jacobi metric}. As we are able to show, the generalized Jacobi metric does inherit the geodesics of the given spacetime. Moreover, our formalism gives a very clear relationship between the constants of motion of the Lorentzian metric and the ones in the generalized Jacobi metric.\\

It is well known that geodesics in a spacetime can be deduced from a variational principle \cite{Wald:1984rg}, where the functional which is  extremized is constructed using the Lagrangian $\mathcal{L}$, which in turn is a function of a Lorentzian metric $g_{\mu\nu}$ .

\begin{equation}\label{lag}
\mathcal{L}=\frac{1}{2}g_{\mu\nu}\frac{dx^{\mu}}{d\lambda}\frac{dx^{\nu}}{d\lambda}=\frac{1}{2}g_{\mu\nu}\dot{x}^{\mu}\dot{x}^{\nu},
\end{equation}

The Hamiltonian of the system is given by the Lengendre transform

\begin{equation}\label{ham}
\mathcal{H}=\sum_{\mu}p_{\mu}\dot{x}^{\mu}-\mathcal{L},
\end{equation}

where

\begin{equation}\label{cor}
p_{\mu}=\frac{\partial \mathcal{L}}{\partial\dot{x}^{\mu}}=\dot{x}_{\mu}
\end{equation}
It is known that Hamilton-Jacobi equation can be written \cite{Wald:1984rg}

\begin{equation}\label{act}
\frac{\partial S}{\partial \lambda}+\frac{1}{2}g^{\mu\nu}p_{\mu}p_{\nu}=0,
\end{equation}
with

\begin{equation}\label{mom}
p_{\mu}=\frac{\partial S}{\partial x^{\mu}},
\end{equation}
where $S$ is the action functional that is built with the Langrangian $\mathcal{L}$ and satisfies $\frac{\partial{S}}{\partial x^{\mu}}=\frac{\partial \mathcal{L}}{\partial \dot{x}^{\mu}}$. On the other hand,the norm $\delta$ of a geodesic tangent vector $\dot{x}^{\mu}$ is given by \footnote{When massive particles are considered then $\delta=-m^2$ and in the null case $\delta=0$. By setting $\mathcal{H}=\delta$ we are enforcing that the Hamiltonian remains constant and therefore motions is restricted to surfaces of constant energy}

\begin{equation}\label{lag2}
\delta=g_{\mu\nu}\dot{x}^{\mu}\dot{x}^{\nu}=2\mathcal{L}=2\mathcal{H},
\end{equation}

where for spacelike geodesics we have $\delta>0$, for the timelike geodesics $\delta<0$ and for the null type we have $\delta=0$. Finally, using \eqref{lag2} and \eqref{act} we obtain

\begin{equation}\label{hamja}
-\frac{\delta}{2}+\frac{1}{2}g^{\mu\nu}p_{\mu} p_{\nu}=0.
\end{equation}
where $p_{\mu}$ is given by \eqref{mom}.

Now consider a spacetime with coordinates $x^{u}=(t,x^i), i=1,2,3$, then equation \eqref{hamja} reads:

\begin{equation}
-\frac{\delta}{2}+\frac{1}{2}\left(g^{tt}p_t p_t+g^{ij}p_{i}p_{j}\right)=0.
\end{equation}

Hence, from this Hamilton-Jacobi equation  we get,
\begin{equation}\label{hamiltone}
\left(\frac{1}{\delta-g^{tt}p_{t}p_{t}}\right)g^{ij}p_{i}p_{j}=1.
\end{equation}

We impose a necessary condition for our method, namely 
\begin{equation}\label{cond:2}
J^{ij}p_{i}p_{j}=1.
\end{equation}

Comparing equation \eqref{hamiltone} with equation \eqref{cond:2} we can identify

\begin{equation}
J^{ij}=\left(\frac{1}{\delta-g^{tt}p_{t}p_{t}}\right)g^{ij}, 
\end{equation}

whose inverse metric is given by

\begin{equation}\label{jacobim}
J_{ij}=\left(\delta-g^{tt}p_{t}p_{t}\right)g_{ij}.
\end{equation}

Note that now the metric $g_{ij}$ is a 3-dimensional Riemannian metric. The metric \eqref{jacobim} is the \textit{generalized Jacobi metric} obtained by projecting the metric \eqref{metric1} over the direction defined by the Killing vector $\partial_{t}$.

This procedure can be generalized to include more Killing vectors in the following way. We name the indices that will run over the coordinates that have a Killing vector with capital letters $A,B,C$. For the remaining coordinates (the ones on which there are not Killing vectors) we use $a,b,c$. Thus, if $d$ is the spacetime dimension we will always have $d=p+q$, where $p$ is the number of coordinates used for projection and $q$ the remaining ones. For example, if we use two Killing vectors for projection we have $A,B=1,2$, and  $a,b=3,4$. Using this notation the generalized Jacobi metric can be written
\begin{equation}\label{jacobim2}
J_{ab}=\left(\delta-g^{AB}p_{A}p_{B}\right)g_{ab}.
\end{equation}
The metric $g_{ab}$ is a $2-$dimensional Riemannian metric. Let us analyze the terms inside the conformal factor.\\
In principle the conserved momenta $p_{A}$ can be any function depending on time, and therefore it could happen that \eqref{jacobim2} represents a family of Riemannian metrics parametrized by $t$. However, if the Lorentzian metric has a Killing vector in the direction of the coordinate $x^{A}$ then, the Lie derivative of the metric in the direction of that Killing vector vanishes. In this particular case, we have that

\begin{equation}
p_{A}=C,
\end{equation}

where $C$ stands for the constant of motion associated with the Killing vector in the direction of the coordinate $x^{A}$. \\
For static spacetimes the constant of motion is the energy $E$, which is associated to the time direction Killing vector $\partial_{t}$, then for the spacetime metrics \eqref{metric1} we have that $\mathcal{L}=\frac{1}{2}(-V^2\dot{t}^2+g_{ij}\dot{x}^{i}\dot{x}^{j})$, then
\begin{equation}\label{killinge}
p_{t}=\frac{\partial \mathcal{L}}{\partial \dot{t}}=-V^2\dot{t}=E.
\end{equation}
Using $g^{tt}=-V^{-2}$  and replacing \eqref{killinge} in \eqref{jacobim} we obtain

\begin{equation}
J_{ij}=\left(V(x)^{-2}E^2+\delta\right)g_{ij},
\end{equation}
where $i,j=r,\theta,\phi$. The previous equation coincides with the result obtained in \cite{Gibbons:2015qja} when we set $\delta=-m^2$, which in its turn corresponds to timelike geodesics.  What we have done removes the temporal part of the metric by considering paths of constant energy, with the advantage that now the remaining metric $g_{ij}$ is a Riemannian metric. 

The generalized Jacobi metric can be extended to time dependent spacetime metrics. We make the metric \eqref{metric1} time dependent by setting $V=V(x,t)$, but now we do not have a Killing vector in the time direction, then the equation \eqref{jacobim} leads to

\begin{equation}
J_{ij}=\left(V^{2}(x,t)\dot{t}^2+\delta\right)g_{ij}.
\end{equation}

where we have used  $p_{t}=-V^2(x,t)\dot{t}$.
The previous result is in agreement with the Jacobi metric\footnote{In \cite{Chanda:2016aph}, due to the fact that there are no surfaces of constant energy, the authors rely on the use of the Eisenhart-Duval lift. In this lift, a dummy variable  which helps to build  5-dimensional spacetime is introduced. On this new spacetime a surface of constant energy can be defined. All of this is possible because the new metric  has a new Killing vector in the direction of the dummy variable, therefore it has a conserved quantity. Thus, we only need to set $\delta=-m^2$, $q_{t}=-m\dot{t}$ and $p_{t}=\frac{\partial \mathcal{L}}{\partial \dot{t}}$. Here $\sigma$ denotes the dummy variable that has been introduced by the Eisenhart-Duval lift. See \cite{Duenas-Vidal:2022kcx} for a detailed discussion.} obtained in \cite{Chanda:2016aph}.

We have no restriction in the type of metric or the number of components, therefore the method can be used for any metric in any dimension. We can expand the expression of the conformal factor in \eqref{jacobim} as much as the number of constants of motion\footnote{Even if there is not a Killing vector in that direction}.

\subsection{\label{sec:2}Static, spherically symmetric spacetimes}
We start by considering a static, spherically symmetric spacetime metric written in spherical coordinates $r>0,\theta\in [0,\pi],\phi\in[0,2\pi)$:
\begin{equation}\label{metric2}
ds^2=g_{tt}(r)dt^2+g_{rr}(r)dr^2+g_{\theta\theta}(r)d\theta^2+g_{\phi\phi}(r)d\phi^2,
\end{equation}

The metric \eqref{metric2} has two Killing vectors $\partial_{t}$ and $\partial_{\phi}$. We first calculate the derivatives of the action $S$:
\begin{eqnarray}
\frac{\partial S}{\partial t}&=&g_{tt}\dot{t}=E\label{energy2d}\\
\frac{\partial S}{\partial \phi}&=&g_{\phi\phi}\dot{\phi}=L\\
\frac{\partial S}{\partial r}&=&g_{rr}\dot{r}.
\end{eqnarray}

If we project over the direction of the vector $\partial_{t}$ we get a $3-$dimensional Riemannian metric:

\begin{equation}\label{jacobim23d}
J_{ij}=\left(\delta-g^{tt} E^2\right)g_{ij},
\end{equation}
where $i,j=r,\theta,\phi$, but if we project over the directions of the Killing vectors $\partial_{t}$ and  $\partial_{\phi}$ the generalized Jacobi metric is
\begin{equation}\label{jacobim22d}
J_{ab}^{2d}=\left(\delta-g^{tt} E^2-g^{\phi\phi}L^2\right)g_{ab}^{2d},
\end{equation}
where $a,b=r,\theta$, and now the result is a two dimensional Riemannian metric.\\
Here we have to clarify something very important regarding the coordinates of the metric $g_{ab}^{2d}$. The components of the $3-$dimensional metric $g_{ij}$ in \eqref{jacobim23d} are the ones corresponding to $r$  , $\theta$ and $\phi$ in the original spacetime metric and because of spherical symmetry we can set\footnote{Note that because of spherical symmetry this setting corresponds to a new projection.} $\theta=\frac{\pi}{2}$, then we obtain a two dimensional metric whose components correspond to $r$ and $\phi$ in the original metric.

We can go farther, although there is not a Killing vector in the direction $\partial_{r}$ we can include the term $g^{rr}p_{r}p_r$ in the conformal factor. Now we set $g^{tt}=-\frac{1}{f(r)}, g^{rr}=g(r), g^{\phi\phi}=\frac{1}{r^2}, g_{\theta\theta}=r^2$ and replace in \eqref{jacobim22d}, we get

\begin{equation}\label{jacobimoned}
J^{1d}=\frac{1}{f(r)}\left(E^2-f(r)\left(\frac{L^2}{r^2}-\delta\right)-\frac{f(r)}{g(r)}\dot{r}^2\right)r^2.
\end{equation} 

Setting the expression which is inside the parentheses in  \eqref{jacobimoned} to zero we get
\begin{equation}\label{radial}
\dot{r}^{2}=\frac{g(r)}{f(r)}\left(E^2-f(r)\left(\frac{L^2}{r^2}-\delta\right)\right),
\end{equation}
The equation \eqref{radial} is a geodesic equation for the radial coordinate. The conformal factor of the generalized Jacobi metric contains the information regarding the geodesics of the metric. It is clear that we do not need more information than the existence of Killing vectors. We will demonstrate that it is not a coincidence but it is intrinsic to our method.\\
Note that the equation \eqref{radial} can be written as 
\begin{equation}
\dot{r}^2=J,
\end{equation}
where the effective potential $J$ is 
\begin{equation}\label{potential}
J=\frac{g(r)}{f(r)}\left(E^2-f(r)\left(\frac{L^2}{r^2}-\delta\right)\right).
\end{equation}

The right side of equation \eqref{jacobimoned} is a function obtained from the intersection of a $3-$ dimensional metric and a $2-$dimensional space.
In the next section we are going to use this $1-$dimensional Jacobi metric\footnote{This $1-$dimensional metric corresponds to an effective potential $V_{eff}$.}for studying the stability of circular orbits.

\subsubsection{Circular geodesics stability}\label{stability1}
It is known that the study of null geodesics stability provides a criteria for the violation of Cosmic Censorship Conjecture when the metric has a Cauchy horizon\cite{Cardoso:2008bp}, where the $1-$ dimensional generalized Jacobi metric \eqref{jacobimoned} plays an essential role. Indeed, enforcing the conditions $J=0,\,\,\,J'=0$ with $\delta=0$ in \eqref{potential} we obtain
\begin{eqnarray}
\frac{E^2}{L^2}&=\frac{f(r)}{r^2},\\
2f(r)&=rf'(r).
\end{eqnarray}

Using $J''$ on the null circular orbit it can be shown by a direct calculation that the Lyapunov exponent $\lambda$ that yields the decay (growing) rate of that orbit is given by \cite{Cardoso:2008bp}
\begin{equation}\label{lambda}
\lambda=\sqrt{\frac{J''}{2 \dot{t}^2}}\bigg\rvert_{r=r_{ph}},
\end{equation}
where $r_{ph}$ stands for the radius of a circular null geodesic.
On the other side, the imaginary part of the quasi-normal mode frequencies of a black hole in the eikonal limit reads \cite{Cardoso:2008bp,Rahman:2018oso}
\begin{equation}
\Im{(\omega)}=-\left(\nu+\frac{1}{2}\right)\lambda,
\end{equation}
where $\nu=0,1,2,...$ is the overtone number. It is clear that the lowest overtone number leads to 
\begin{equation}
\Im(\omega)_{min}=-\frac{\lambda}{2}.
\end{equation}
Moreover, if the spacetime has a Cauchy horizon we can calculate its surface gravity, then a coefficient $\beta_{ph}$ is defined as:
\begin{equation}
\beta_{ph}=-\frac{\Im(\omega)_{min}}{\kappa_{-}}=\frac{\lambda}{2\kappa_{-}}>\frac{1}{2},
\end{equation}
where $\kappa_{-}$ is the surface gravity of the Cauchy horizon. The factor $\beta_{ph}$ is calculated in the eikonal limit, and therefore it is defined for the photon sphere circular null orbits. If $\beta_{ph}$ is bigger than $1/2$ then the Strong Cosmic Censorship is violated \cite{Cardoso:2008bp}. 

Using the equation  \eqref{metric2} and \eqref{energy2d} together with \eqref{lambda} we find an expression for $\lambda$ \cite{Cardoso:2008bp,Rahman:2018oso}
\begin{equation}\label{lambda2}
\lambda=\sqrt{\frac{g(r_{ph})}{2}\left(\frac{2f(r_{ph})}{r^2_{ph}}-f''(r_{ph})\right)}.
\end{equation}
Then, with $\kappa_{-}=\frac{1}{2}g'(r_{-})$ the result for $\beta_{ph}$ con be written
\begin{equation}
\beta_{ph}=\sqrt{\frac{g(r_{ph})}{2 (g'(r_{-}))^2}\left(\frac{2f(r_{ph})}{r^2_{ph}}-f''(r_{ph})\right)}.
\end{equation}

When considering stationary spacetimes the problem of defining a Jacobi metric becomes difficult. The different Jacobi metrics proposed in \cite{Gibbons:2015qja} and \cite{Chanda:2019guf} for the same metric are different. Our new technique provides a unique way of calculating a generalized Jacobi metric for static, time-dependent and stationary spacetimes.

In the following section we show the results obtained for stationary axisymmetric spacetimes, then we particularize for Kerr spacetime.

\subsection{\label{sec:3} Stationary spacetimes}

Using the equation \eqref{jacobim2} we proceed to calculate the generalized Jacobi metric for stationary axisymmetric spacetimes\footnote{We focus in the particular case when the surfaces generated by the Killing vectors admit orthogonal surfaces, this property is known as orthogonal transitivity. For stationary axisymmetry spacetimes both Killing vectors conmmute. Stationary axisymmetric spacetimes on which the orthogonal property is not satisfied are very rare.}. We have to take into account more terms in the conformal factor. In particular, we will deduce the geodesic equations for the Kerr metric, showing that the geodesic structure of the original metric is inherited by the generalized Jacobi metric. We will also discuss the stability of of circular geodesics.\\

We start by considering the following general metric

\begin{align}\label{stationary}
ds^2=g_{tt}(r)dt^2+2g_{t\phi}(r,\theta)dtd\phi+g_{rr}(r,\theta)dr^2\nonumber\\
+g_{\theta\theta}(r,\theta)d\theta^{2}+g_{\phi\phi}(r,\theta)d\phi^2.
\end{align}

The $3$-dimensional generalized Jacobi metric can be easily calculated. First, we need to calculate $p_{t}$ and $p_{\phi}$:
\begin{eqnarray}
p_{t}&=&g_{tt}\dot{t}+g_{t\phi}\dot{\phi}=E,\label{energy2}\\
p_{\phi}&=&g_{t\phi}\dot{t}+g_{\phi\phi}.\dot{\phi}=L.\label{momentum}
\end{eqnarray}

Solving the previous system we obtain

\begin{eqnarray}
\dot{t}&=&\frac{g_{\phi\phi}E+g_{t\phi}L}{-g^2_{t\phi}+g_{\phi\phi}g_{tt}},\label{tdot}\\
\dot{\phi}&=&\frac{-g_{tt}L-g_{t\phi}E}{-g^2_{t\phi}+g_{\phi\phi}g_{tt}}.\label{phidot}.
\end{eqnarray}

If we only consider the projection over one Killing vector we won't recover the geodesics\footnote{The metric \eqref{jacobim2} is rewritten as

\begin{equation}\label{jacobimest3d}
J_{ij}=\left(\delta-g^{tt}p_{t}p_{t}-g^{\phi\phi}p_{\phi}p_{\phi}\right)g_{ij},
\end{equation}
where now $i,j=r,\theta,\phi$. Using  \eqref{energy2} and \eqref{momentum} in \eqref{jacobimest3d}, the $3-$dimensional generalized Jacobi metric $J_{ij}$ is written
\begin{equation}\label{jacobi3d}
J_{i,j}^{3d}=\left(\delta-\frac{1}{g^2_{t\phi}+g_{\phi\phi}g_{tt}}\left(g_{\phi\phi}E^2-g_{tt}L^2\right)\right)g_{i,j}^{3d}.
\end{equation}
This metric does not inherit the geodesic structure of the metric \eqref{stationary}.}, we need to use both Killing vectors. We won't be able to recover the radial geodesic equation unless we consider an extra term in the conformal factor, the term $g^{\phi\phi}p_{\phi\phi}$. Including this term we obtain

\begin{equation}\label{jacobim3d}
J_{ab}^{2d}=\left(\delta-g^{tt}p_{t}p_{t}-2g^{t\phi}p_{t}p_{\phi}-g^{\phi\phi}p_{\phi}p_{\phi}\right)g_{ab}^{2d},
\end{equation}

whence
\begin{equation}\label{jacobiest2d}
J_{ab}^{2d}=\left(\delta-\frac{1}{-g^2_{t\phi}+g_{\phi\phi}g_{tt}}\left(g_{\phi\phi}E^2-2g_{t\phi}E L+g_{tt}L^2\right)\right)g_{ab}^{2d}.
\end{equation}

where we have used equations \eqref{energy2} and \eqref{momentum}. The indices in $g_{ab}$ run over a $2$-dimensional manifold, $a,b=r,\theta$.\\
Here we can see the particularity of the stationary spacetimes. The conformal factor that multiplies the metric $g_{ab}^{2d}$ includes the energy $E$ and $L$ and there is no way to project only on surfaces of constant energy $E$. There is always a mixed term with the angular momentum. If we project over both Killing vectors $\partial_{\phi}$ and $\partial_{t}$, then the generalized Jacobi metric would be \eqref{jacobiest2d}. The generalized Jacobi metric \eqref{jacobiest2d} solves the problem  of obtaining the geodesics from a Riemannian metric.\\
For stationary spacetimes we will have two different reparametrizations when comparing the equations of motion for $g_{ij}$ and the equations of motion for the generalized Jacobi metric. This is the reason why all previous attempts have failed.

\subsubsection{Circular geodesics stability}
Following the procedure depicted in section \eqref{stability1} we can study the circular orbits stability. We start by finding the radial geodesic equation using the generalized Jacobi metric. In the conformal factor we include the terms associated with both Killing vectors $\partial_t$ and $\partial_{\phi}$, and also  the term $p_{r}=g_{rr}\dot{r}$, then
\begin{equation}\label{jacobi2dr}
J^{1d}=\left(\delta-\frac{1}{-g^2_{t\phi}+g_{\phi\phi}g_{tt}}\left(g_{\phi\phi}E^2-2g_{t\phi}E L+g_{tt}L^2-g_{rr}\dot{r}^2\right)\right)g_{\phi\phi}.
\end{equation}
By equating the conformal factor to zero we arrive to
\begin{equation}
\dot{r}^2=-\frac{1}{g_{rr}}\left(\delta-\frac{1}{-g^2_{t\phi}+g_{\phi\phi}g_{tt}}\left(g_{\phi\phi}E^2-2g_{t\phi}E L+g_{tt}L^2\right)\right)=J.
\end{equation}
By imposing $J=0$ and $\delta=0$ we obtain an expression for $l=\frac{E}{L}$ 
\begin{equation}
l=\frac{g_{t\phi}}{g_{tt}}\pm\sqrt{\left(\frac{g_{t\phi}}{g_{tt}}\right)^2-\frac{g_{\phi\phi}}{g_{tt}}}.
\end{equation}
Together with $J=0$ we impose that $J'=0$, then we get
\begin{equation}
g'_{\phi\phi}-2g'_{t\phi}l+g'_{tt}l^2=0.
\end{equation}
Finally, 
\begin{equation}
J''=\frac{E^2}{g_{rr}}\left(\frac{g''_{\phi\phi}-2g''_{t\phi}l+g''_{tt}l^2}{g_{\phi\phi} g_{tt}-g^2_{t\phi}}\right).
\end{equation}
We obtain the exact same result as presented in \cite{Rahman:2018oso}. In order to find $\lambda$ we use equation \eqref{lambda} together with \eqref{tdot} and find 
\begin{equation}
\lambda=\sqrt{\frac{\left(g_{\phi\phi} g_{tt}-g^2_{t\phi}\right)\left(g''_{\phi\phi}-2g''_{t\phi}l+g''_{tt}l^2\right)}{2g_{rr}\left(g_{\phi\phi}-g_{t\phi}l\right)^2}},
\end{equation}
then, using $l^{2}=\frac{g_{\phi\phi}-2g_{t\phi}l}{g_{tt}}$ we arrive to
\begin{equation}\label{lambdag}
\lambda=\sqrt{\frac{\left((g''_{\phi\phi}g_{tt}-g''_{tt}g_{\phi\phi})-2(g''_{t\phi}g_{tt}-g''_{tt}g_{t\phi})l\right)(-g^2_{t\phi}+g_{\phi\phi}g_{tt})}{2g_{tt}g_{rr}(g_{\phi\phi}+g_{t\phi}l)}}
\end{equation}
The expression \eqref{lambdag} for the Lyapunov exponent is the same as equation (44) in \cite{Rahman:2018oso}. When $g_{t\phi}=0,g_{tt}=f(r),g_{rr}=\frac{1}{g(r)},g_{\phi\phi}=r^2$ we recover \eqref{lambda2},which can be compared with the results in \cite{Rahman:2018oso}. If we want to calculate $\beta_{ph}$ when evaluate everything in $r_{ph}$ which can be found by solving $g_{tt}=0.$\\
We have shown that all the information regarding the circular geodesic stability is encoded in the generalized Jacobi metric. We have taken as an example the Lyapunov exponent of the circular orbit, but there are plenty of examples where the Jacobi metric can be used. In the next section we show how the constants of motion are inherited when the projection is done. 

\section{\label{sec:4} The constants of motion and geodesics}
Let an $n+1$-dimensional spacetime metric be

\begin{equation}\label{metric22}
ds^2=-g_{tt}dt^2+g_{ij}dx^idx^j,
\end{equation}

then, the corresponding Hamilton's equations of motion coming from \eqref{ham} and \eqref{metric22} are:

\begin{eqnarray}
\dot{x}^i&=&\frac{\partial \mathcal{H}}{\partial p_{i}}=\frac{1}{2}g^{ij}p_{j},\label{pos1}\\
\dot{p}_i&=&-\frac{\partial \mathcal{H}}{\partial x^i}=-\frac{1}{2}\frac{\partial g^{ij}}{\partial x^i}p_{i}p_{j}-\frac{\partial Q(x^i)}{\partial x^{i}},\label{mo1}
\end{eqnarray}

where 

\begin{equation}\label{qexp}
Q(x^i)=g^{tt}p_{t}p_{t}+g^{t\phi}p_{t}p_{\phi}
\end{equation}

For static spacetimes like \eqref{metric22}, the term involving $Q(x^{i})$ is given by

\begin{equation}
\frac{\partial Q(x^i)}{\partial x^{i}}=\frac{1}{2}E^2\frac{\partial g^{tt}}{\partial x^i},
\end{equation}

and therefore, when the metric ${g_{ij}}$ is replaced by the Jacobi metric \eqref{jacobim}, the equations \eqref{pos1},\eqref{mo1} transform to

\begin{eqnarray}
\dot{x}^i&=&\frac{\partial \tilde{\mathcal{H}}}{\partial p_{i}}=\frac{g^{ij}p_{j}}{2\left(\delta-g^{tt}E^2\right)}\label{pos}\\
\dot{p}_i&=&-\frac{\partial \tilde{\mathcal{H}}}{\partial x^i}=-\frac{1}{2\left(\delta -g^{tt}E^2\right)}\left(\frac{1}{2}\frac{\partial g^{ij}}{\partial x^i}p_{i}p_{j}+\frac{1}{2}E^2\frac{\partial g^{tt}}{\partial x^i}\right)\label{mo}.
\end{eqnarray}

We can see that the pair of equations \eqref{pos},\eqref{mo} are a reparametrization of the equations \eqref{pos1},\eqref{mo1}, where the reparametrization\footnote{Here we assume that the momentum $p_{\mu}$ given in equation \eqref{cor} is equal to $\tilde{p}_{i}=\frac{1}{2}J_{ij}{\dot{\tilde{x}}^{j}}$, the momentum associated to the generalized Jacobi metric over $\mu=i$. From this equality  we can obtain the reparametrization \eqref{rep:2} that allows us to obtain the geodesics for the generalized Jacobi metric associated with \eqref{metric1}. For stationary metrics, due to the presence of the term $g^{t\phi}$ the momenta are not the same and therefore  there is not a unique reparametrization. This is another reason why the previous attempts to build a Jacobi metric for a stationary spacetimes have trouble. Our proposal provides the correct parametrization.}  is given by

\begin{equation}
d\tilde{s}=\frac{1}{\left(\delta-g^{tt}E^2\right)}ds,
\end{equation}
where $\tilde{s}$  is the proper time for \eqref{pos1} and \eqref{mo1}, and  
$s$ is the proper time for \eqref{pos} and \eqref{mo}.

It is straightforward to show that if there is a constant of motion $K$  then $\{K,\mathcal{H}\}=0$ implies that $\{K,\tilde{\mathcal{H}}\}=0$, where $\{.,.\}$ represents the Poissson bracket \cite{Tsiganov:2001,Chanda:2016aph}. The constants of motion calculated for the spatial part of the metric are still constants of motion for the Jacobi metric.\\

For stationary spacetimes, we must consider, in the equation \eqref{qexp}, the whole expression involving the energy  $E$ and angular momentum $L$ as

\begin{equation}
Q(x^i)=g^{tt}p_{t}p_{t}+g^{t\phi}p_{t}p_{\phi}+g^{\phi\phi}p_{\phi}p_{\phi},
\end{equation}

then, 

\begin{equation}
Q(x^{i})=g^{tt}E^2+g^{t\phi}E L+g^{\phi\phi}L^2.
\end{equation}

The corresponding reparametrization is given by
\begin{equation}\label{rep:2}
\frac{d \sigma}{d\tilde{s}}=\left(\delta+\frac{g_{\phi\phi}E^2+2g_{t\phi}EL}{g_{\phi\phi} g_{tt}+g_{t\phi}^2}-g^{\phi\phi}L^2\right),
\end{equation}

where now $\sigma$ is the proper time for \eqref{pos} and \eqref{mo}.

This requirement tells us that we cannot project over surfaces of constant energy without considering a surface defined by constant angular momentum also. Therefore, the corresponding Jacobi metric \eqref{jacobim} will be a $2$-dimensional metric. We have shown that the equations of motion of the Lorentzian spacetime metric are also equations of motion of the generalized Jacobi metric. Here we show that the only possible reparametrization involves the conformal factor of the generalized Jacobi metric. In the followng section we  particularize the method for Kerr spacetimes.

\section{\label{sec:5}The generalized Jacobi metric for Kerr spacetime}
The Kerr spacetime line element in Boyer-Linquidst coordinates is written

\begin{align}\label{kerrm}
ds^2=-\left(1-\frac{2Mr}{\Sigma}\right)dt^2-\frac{4aMr\sin^2(\theta)}{\Sigma}dtd\phi+\frac{\Sigma}{\Delta}dr^2\nonumber\\
+\Sigma d\theta^2+\frac{\Upsilon}{\Sigma}\sin^2(\theta)d\phi^2,
\end{align}

where

\begin{eqnarray}
\Sigma&=&r^2+a^2\cos^2(\theta)\label{sigma},\\
\Delta&=&r^2+a^2-2Mr\label{delta},\\
\Upsilon&=&(r^2+a^2)^2-a^2\Delta \sin^2(\theta)\label{upsilon}.
\end{eqnarray}

The Kerr metric has two killing vectors $\partial_t$ and $\partial_\phi$ commuting with each other. Both of them satisfy the killing equation\footnote{Moreover, the Kerr metric has a Killing tensor $\sigma_{\mu\nu}$ that satisfies a Killing equation. This tensor cannot be written as a tensor product of the Killing vectors $\partial_t$ and $\partial_\phi$.}. As we have seen before, for stationary spacetimes the energy $E$ and angular momentum $L$ are linear combinations of $\dot{t} $ and $\dot{\phi}$ 

\begin{eqnarray}
	p_{t}&=&-\left(1-\frac{2Mr}{\Sigma} \right) \dot{t}-\frac{2aMr\sin^2(\theta)}{\Sigma}\dot{\phi}=E,\label{energy}\\
	p_{\phi}&=&-\frac{2aMr\sin^2(\theta)}{\Sigma}\dot{t}+\frac{\Upsilon}{\Sigma}\sin^2(\theta)\dot{\phi}=L.
\end{eqnarray}

By replacing the components of the metric \eqref{kerrm} in the metric \eqref{jacobim3d} we obtain a $3-$dimensional metric\footnote{If we consider the projection over the Killing vector $\partial_{t}$ only we obtain a $3-$dimensional metric: 

\begin{equation}
J_{ab}=\left(\delta+\frac{\Upsilon E^2+4aMrEL_z}{\Sigma \Delta}\right)g_{ab}.
\end{equation} 

This Jacobi metric is different from the ones obtained in \cite{Gibbons:2015qja,Chanda:2016aph}. These metrics, although they inherit information from the spacetime metric, they do not inherit all of the information regarding the geodesics. Our result shows something very interesting, we cannot have a Jacobi metric in three dimensions that inherits the geodesic structure of the Kerr spacetime metric.}. For stationary metrics, we have to go below $3-$dimensions.Using \eqref{jacobiest2d} we get

\begin{equation}\label{Kerr2d}
J_{ab}=\left(\delta+\frac{\Upsilon E^2+4aMrEL}{\Sigma\Delta}-\left(\frac{\Delta-a^2\sin^{2}(\theta)}{\Sigma \Delta\sin^2(\theta)}\right)L_z^2\right)g_{ab},
\end{equation}

where $a,b=r,\theta$ and therefore 

\begin{align}
g_{ab}dx^{a}dx^{b}=\left(\frac{\Sigma}{\Delta}dr^2+\Sigma d\theta^2\right).
\end{align} 

Dividing the line element $ds^{2}=J_{ab}dx^{a}dx^{b}$ by $ds^2$, replacing $\Upsilon$ defined in \eqref{upsilon} and rearranging terms we obtain

\begin{align}
r^2\delta+\frac{1}{\Delta}\left((r^2+a^2)E+aL\right)^2-\frac{\Sigma^2}{\Delta}\dot{r}^2=\nonumber\\
\frac{1}{\sin^{2}(\theta)}\left(a\sin^{2}\left(\theta\right)E+L_z\right)^2+\Sigma^2\dot{\theta}^2-\delta a^2\cos^2(\theta).
\end{align}

Clearly, since the left side of the equation does not depend on $\theta$ and the right side does not depend on $r$ then the previous equation is separable in $r$ and $\theta$ , therefore

\begin{eqnarray}
r^2\delta+\frac{1}{\Delta}\left((r^2+a^2)E+aL_z\right)^2-\frac{\Sigma^2}{\Delta}\dot{r}^2=K\label{Keq:1},\\
\Sigma^2\dot{\theta}^2+\frac{1}{\sin^{2}(\theta)}\left(a\sin^2(\theta)E+L_z\right)^2-\delta a^2\cos^2(\theta)=K,\label{Keq:2}
\end{eqnarray}

where $K$ is a constant. This constant is known as the Crater constant and it is not associated to any symmetry of the Kerr metric.

Finally, equations \eqref{Keq:1} ,\eqref{Keq:2} can be written

\begin{eqnarray}
\Sigma^2\dot{r}^2&=&\mathcal{R},\\
\Sigma^{2}\dot{\theta}^2&=&\Theta,
\end{eqnarray}

where

\begin{eqnarray}
\mathcal{R}&=&\left((r^2+a^2)E+aL_z\right)^2-\Delta\left(K-r^2\delta\right)\\
\Theta &=& K-\frac{1}{\sin^2{\theta}}\left(a\sin^{2}(\theta)E+L_z\right)^2+\delta a^2\cos^2(\theta).
\end{eqnarray}

The  remaining pair of geodesic equations can be obtained from \eqref{tdot} and \eqref{phidot}, thus

\begin{eqnarray}
\Sigma\dot{t}&=&-\frac{1}{\Delta}\left(E \Upsilon +2amrL_z\right)\\
\Sigma\dot{\phi}&=&-\frac{1}{\Delta}\left(2amrE-\left(\Sigma-2mr\right)\frac{L_z}{\sin^{2}(\theta)}\right)
\end{eqnarray}

We have found a $2$-dimensional Jacobi metric that inherits the geodesics of the Kerr spacetime. In the next section we will analyze some geometric properties of the $2-$dimensional generalized Jacobi metric \eqref{jacobim22d}.

\section{\label{sec:6}About the curvatures and black hole spacetimes}
In section \ref{sec:3} a generalized Jacobi metric for stationary spacetimes was calculated. Here we calculate the reduced generalized  Jacobi metric for asymptotically flat, static (stationary), spherically symmetric  spacetimes. 
In general, the Jacobi metric has a conformal factor multiplying an \textit{optical} metric, for example, in the metric \eqref{jacobim22d} the conformal factor is $ \left(\delta-g_{tt} E^2-g_{\phi\phi}L^2\right)$ and an optical metric is $g_{ab}^{2d}$. The optical metric can be obtained by setting $ds^2=0$ and solving for $dt$. In the case of the metric \eqref{metric2} the optical metric is
\begin{equation}\label{opticm}
dt^2=\frac{1}{g_{tt}}\left( g_{rr}dr^2+r^2d\phi^2\right).
\end{equation}

The geodesic curvature of the metric \eqref{opticm} is given by
\begin{equation}
k_{g}=-\frac{1}{2}\sqrt{\frac{1}{g_{tt}g_{rr}}}\left(\frac{2g_{tt}-rg'_{tt}}{2r}\right)
\end{equation}

Due to the fact that the geodesic curvature vanishes for null circular geodesics we can find them by solving $k_{g}=0$. This leads to $2g_{tt}=rg'_{tt}$, which enforces a condition over the form of the factor $g_{tt}$. When calculating the generalized Jacobi metric  the optical metric should appears when we set $\delta=0$. It indeed appear, up to a factor, when we set $\theta=\pi/2$ in the metric \eqref{jacobim23d}. Therefore, a calculation of the Gaussian and geodesic curvatures can be carried out. However, if we consider directly the two dimensional metric \eqref{jacobim22d} the calculation turn out to be different. The coordinates that parametrize the $2-$dimensional generalized Jacobi metric are $r,\theta$ instead of $r,\phi$. Moreover the conformal factor of the metric \eqref{jacobim22d} is a function of $r$ and $\theta$. Even in the case were we can calculate the Gaussian curvature it will be a complicated expression. The geodesic curvature is the component of acceleration in the direction normal to the geodesic. Then, the points where $\kappa_g=0$ could be identified.

In the case of massive particles trajectories we can calculate the geodesic curvature of \eqref{jacobim23d}, then after setting $\theta=\pi/2$ the geodesic curvature reads
\begin{equation}\label{gedesicc}
k_{g}=\frac{2g_{tt}\left((E/\delta)^2-g_{tt}\right)-r(E/\delta)^2g'_{tt}}{2mr\sqrt{g_{tt}g_{rr}}((E/\delta)^2-g_{tt})^{3/2}}.
\end{equation}
Solving the equation $k_{g}=0$ for $E/\delta$ we obtain
\begin{equation}
\left(\frac{E}{\delta}\right)^2=-\frac{2g_{tt}}{2g_{tt}-rg'_{tt}}.
\end{equation}

Due to the fact that the left hand side of the previous equation is positive, the only possible way to have null geodesics is by enforcing the restriction $2g_{tt}-rg'_{tt}<0$.\\
The stability of the radial geodesics is determined by the sign of the Gaussian curvature, therefore using \eqref{gedesicc} and the Gaussian curvature of the generalized Jacobi metric the circular orbits stability can be studied\footnote{The Gaussian curvature $K_{G}$ of a two dimensional metric satisfies 
\begin{equation}
K_{G}=\frac{1}{2}R,
\end{equation}

where $R$ is the scalar curvature of the $2$-dimensional generalized Jacobi metric.}. This is related  to the previous sections where we have calculated the $J$ obtained from the generalized Jacobi metric procedure. Since we have shown that the geodesics are stored in the generalized Jacobi metric we expect that there is a geometric way of studying its stability. The simplest case is studied in \cite{Cunha:2022nyw}, where the Gaussian curvature and the geodesic curvature of a Jacobi metric are used for studying the stability of circular geodesics. We can try a similar procedure but with the generalized Jacobi metric, it will allow us to study stationary spacetimes and other much more complicated  types of metrics.  We left  this for future work.

In the following we present the $2$-dimensional generalized Jacobi metric for different black hole spacetimes, we can see that all of them have a similar dependence on both coordinates $r,\theta$. Moreover, all of them have the coordinate $\theta$ inside the conformal factor. These geometries are going to be difficult to study because of this dependence, but the tools of the Riemannian geometry can be applied safely. Let us show the results for the Gaussian curvature of the reduced Jacobi metric for different black hole metrics. 

For a static, spherically symmetric spacetimes of the form

\begin{equation}
ds^2=-f(r)dt^2+\frac{1}{f(r)}dr^2+d\Omega^2_{2},
\end{equation}

where $d\Omega^2_{2}$ is the $S^{2}$ sphere element in spherical coordinates, we have that the $2$-dimensional Jacobi metric \eqref{jacobim22d} is given by

\begin{equation}\label{2dJac}
J^{2d}_{ij}dx^{i}dx^{j}=\left(\delta+\frac{E^2}{f(r)}-\frac{L^2}{r^2\sin^2(\theta)}\right)\left(\frac{dr^2}{f(r)}+r^2d\theta^2\right). 
\end{equation}

The previous metric has a Ricci scalar, and therefore a Gaussian curvature, depending on the first and second derivatives of the function $f(r)$. The Gaussian curvature result is shown in Appendix \eqref{app:1}. Due to the dependence on the $\theta$ variable the Gaussian curvature is going to be a complicated expression. Here we particularize for the null case $\delta=0$, then we have
\begin{equation}\label{gaussianG}
\begin{split}
K_{G}^{\delta=0}&=\frac{Q_G}{4 \left(E^2 r^2-L^2 f(r) \csc ^2(\theta )\right)^3}
 \end{split}
\end{equation}

where
\begin{equation}
\begin{split}
Q_G&=-E^4 r^6 f'(r)^2\\
&-2 E^2 r^2 f(r)^2 \left(L^2 r^2 \csc   ^2(\theta ) f''(r)+4 L^2 r \csc ^2(\theta ) f'(r)-2 L^2 (\cos (2\theta )+2)\csc^4(\theta )\right)\\
&+4 L^2 f(r)^3 \csc^2(\theta )\left(2 E^2 r^2-L^2 \csc ^4(\theta )\right)+f(r) \left(2
   E^4 r^6 f''(r)+3 E^2 L^2 r^4 \csc^2(\theta )
   f'(r)^2\right)
   \end{split}
\end{equation}
In the following we show how to  find the radial geodesics only by using the generalized Jacobi metric and a reparametrization of the proper time.
\subsubsection*{The Schwarzschild metric}
We start with the Schwarzschild metric 

\begin{equation}
ds^2=-\left(1-\frac{2M}{r}\right)dt^2+\frac{dr^2}{\left(1-\frac{2M}{r}\right)}+r^2d\theta^2+r^2\sin^2(\theta)d\phi^2,
\end{equation}

where the event horizon is located at $r_{S}=2M$. Using the equation \eqref{2dJac} we can calculate a $2$-dimensional Jacobi metric. Due to the fact that the Schwarzschild metric has two  Killing vectors  we know that we can associate two conserved quantities, the energy $E$ and the momentum $L$ in four dimensions. Therefore, we can project over surfaces of constant $E$ and constant $L$. Thus, our generalized $2$-dimensional Jacobi metric line element is given by

\begin{align}\label{jacobis2}
J^{2d}_{ij}dx^{i}dx^{j}=\left(\delta-\frac{E^2}{\frac{2M}{r}-1}-\frac{L^2}{r^2\sin^2(\theta)}\right)d\Omega^{2}_{S},
\end{align}
where
\begin{equation}
d\Omega^{2}_{S}=\left(\frac{dr^2}{1-\frac{2M}{r}}+r^2d\theta^2\right).
\end{equation}

Then, we can calculate the Gaussian curvature of the metric \eqref{jacobis2}. See appendix \ref{app:1}. We can use \eqref{gaussianG} for calculating the Gaussian curvature of the null ($\delta=0$) generalized Jacobi metric \eqref{jacobis2},  we get
\begin{equation}\label{jacobimSd0}
K_{G}^{\delta=0}=\frac{Q_{G}}{\left(E^2 r^3+L^2 \csc ^2(\theta ) (2 M-r)\right)^3},
\end{equation}

where
\begin{equation}
\begin{split}
Q_G&=2 E^4 M r^5 (3 M-2 r)\\
&+3 E^2 L^2 r^2 \csc ^4(\theta )
(2 M-r) \left(-5 M^2+M \cos (2 \theta ) (5 M-2 r)+6 M r-2
r^2\right)\\
&+2 L^4 \csc ^6(\theta ) (2 M-r)^3.
\end{split}
\end{equation}

At the horizon the Gaussian curvature \eqref{jacobimSd0} becomes
\begin{equation}
\lim_{r\rightarrow 2M}K_{G}^{\delta=0}=-\frac{1}{8E^2 M^2}.
\end{equation}

Similarly, when $r\rightarrow \infty$ we have that $K_{G}^{\delta=0}\rightarrow 0$. This is the exact same result obtained with the Gaussian curvature of the Jacobi metric (setting $\theta=\pi/2$) obtained projecting the Schwarzschild metric calculated in \cite{Gibbons:2015qja}.   This  leads us to conclude that the Gaussian curvature is growing from negative values to zero in both, the Jacobi metric and the generalized Jacobi metric introduced in this article. Our generalized Jacobi metric stores the same information regarding the intrinsic curvature at the horizon and at infinity.  

\subsubsection*{The Reissner-Nordström metric}
The spacetime metrics that have two  Killing vectors can be reduced to a 2-dimensional Jacobi metric. The  Reissner-Nordström is a natural generalization of the Schwarzschild metric by including electrical charge. This metric is written as

\begin{align}
ds^2=-\left(1-\frac{2M}{r}-\frac{Q^2}{r^2}\right)dt^2+
\frac{dr^2}{\left(1-\frac{2M}{r}-\frac{Q^2}{r^2}\right)}\nonumber\\+r^2d\theta^2+r^2\sin^2(\theta)d\phi^2.
\end{align}

The generalized 2-dimensional Jacobi metric coming from the previous metric is given by

\begin{align}\label{jacobir}
J^{2D}_{ij}dx^{i}dx^{j}=\left(\delta r^2+\frac{E^2r^4}{Q^2-2Mr+r^2}-\frac{L^2}{\sin^2(\theta)}\right)d\Omega^{2}_{R-N},
\end{align}
where
\begin{equation}
d\Omega^{2}_{R-N}=\left(\frac{dr^2}{Q^2-2Mr+r^2}+r^2d\theta^2\right).
\end{equation}

As before, we can calculate the Gaussian curvature of the reduced generalized Jacobi metric \eqref{jacobir}. We will focus in the null case of the Gaussian curvature calculated in \eqref{gaussianG}. For simplicity we take $M=0$. Thus, we use equation \eqref{gaussianG} with $f(r)=1-\frac{Q^2}{r}$ and calculate the Gaussian curvature of the result, we obtain

\begin{equation}
\begin{split}
K_G=\frac{Z_{G}}{2\left(\text{EE}^2 r^4+L^2
   \csc ^2(\theta ) (Q-r) (Q+r)\right)^3},
\end{split}
\end{equation}
where 

\begin{equation}\label{jacobimRNd0}
\begin{split}
Z_{G}&=E^4 Q^2 r^6 \left(2 Q^2-3 r^2\right)\\
&+\frac{3}{2}E^2 L^2 r^2 \csc ^4(\theta ) (r-Q) (Q+r) \left(2 Q^4+Q^2 \cos (2 \theta ) \left(r^2-2 Q^2\right)-3 Q^2 r^2+2 r^4\right)\\
&+L^4\csc ^6(\theta ) \left(Q^2-r^2\right)^3.
\end{split}
\end{equation}

In the asymptotic limit $r\rightarrow \infty$ the Gaussian curvature defined in \eqref{jacobimRNd0} goes to zero as expected, and in the near horizon limit we have that
\begin{equation}
K_{G}^{\delta=0}=-\frac{1}{E^2 Q^2}.
\end{equation}

The previous result is the same for the Jacobi metric calculated  in \cite{Das:2016opi}. The generalized Jacobi metric defined in this article behaves as expected in the near horizon limit and asymptotically. The previous calculations can be done for the Reissner-Nördstrom metric with $M\neq 0$ and the results are in accordance.

\subsubsection*{The Kerr-Newman metric}
The Kerr-Newman metric in Boyer Linquidst coordinates is

\begin{align}\label{KN}
ds^2=-\left(1-\frac{2Mr-Q^2}{\Sigma}\right)dt^2\nonumber \\-\frac{2a\sin^2(\theta)(2Mr-Q^2)}{\Sigma}dtd\phi
+\frac{\Sigma}{\tilde{\Delta}}dr^2+\Sigma d\theta^2\nonumber \\+\frac{\Upsilon}{\Sigma}\sin^2(\theta)d\phi^2,
\end{align}

where

\begin{equation}
\tilde{\Delta}=r^2+a^2-2Mr+Q^2,
\end{equation}

with $\Sigma$ and $\Upsilon$ given in \eqref{sigma} and \eqref{upsilon} respectively.\\
The generalized Jacobi metric for the Kerr-Newman metric \eqref{KN} is

\begin{align}
&J_{ij}^{2d}dx^{i}dx^{j}=\\ &\left(\delta+\frac{\Upsilon E^2+2(2aMr+Q)EL-\frac{L^2\tilde{\Delta}}{\sin^2(\theta)}+a^2L^2}{\Sigma\tilde{\Delta}}\right) d\Omega^{2}_{KN},
\end{align}

where
\begin{equation}
d\Omega^{2}_{KN}=\left(\frac{\Sigma}{\tilde{\Delta}}dr^2+\Sigma d\theta^2\right).
\end{equation}

As in the all other examples the Gaussian curvature can be calculated using \eqref{gaussianG}. The result is a complicated expression. Here we present the near horizon limit for the case $Q=0$, namely the Kerr metric:
\begin{equation}\label{jacobigK}
\lim_{r\rightarrow M+\sqrt{M^2-a^2}}K_{G}^{\delta=0}=\frac{P_{G}}{a^2 \left(4 E M^2 (a E+L)+a L^2\right)^2},
\end{equation}
where 
\begin{equation}
P_{G}=2 (a^2-M^2) \left(a^2 \left(L^2-4 E^2 M^2\right)+8 E^2 M^3 \left(\sqrt{M^2-a^2}+M\right)+4 a E L M
   \left(\sqrt{M^2-a^2}+M\right)\right),
\end{equation}

and we have set $\theta=\pi/2$. When $a=0$ we recover the Schwarzschild result \eqref{jacobimSd0}. 
It is important to note that in all previous cases the resulting generalized Jacobi metric depends on both coordinates $r, \theta$, therefore the calculations are going to be really cumbersome. This Gaussian curvature tends to zero at infinity as expected.

\section{\label{discussion} Discussion and final remarks}

We have presented the \textit{generalized Jacobi metric}. This metric encodes some properties of the original spacetime, but the most important fact is that it inherits its null and timelike geodesics, therefore we are able to recover the geodesic equations for all variables. Until now, the Jacobi metric described in the literature worked very well for static, asymptotically flat, spherically symmetric spacetimes \cite{Gibbons:2015qja,Chanda:2016aph}. The method has been widely used for studying the dynamics of particles moving in those spacetimes by using the tools of Riemannian geometry \cite{Chanda:2016sjg,Das:2016opi,Arganaraz:2019fup}. However, the method has some difficulties when applied to stationary spacetimes . The most  important of them was that the geodesic structure of spacetime was not inherited completely \cite{Chanda:2019guf}. 

Due to the fact that in the stationary case the usual formalism does not work well, we consider that the only possible approach is to use a different definition of Jacobi metric. The method proposed here solves the aforementioned problems and reduces to the already known results for the static and dynamic cases. For stationary spacetimes we have to go below three dimensions. When we project over surfaces defined by two Killing vectors, then a $2-$dimensional generalized Jacobi metric can be calculated, and from this metric we were able to recover all the geodesic structure.
Due to the fact that the geodesic structure is inherited  by the generalized Jacobi metric the stability of null circular orbits can be inferred from it. We have shown how to calculate the Lyapunov exponent of the orbits obtaining the same results  as in the literature. We discussed the application of this to the study of the violation of the Strong Cosmic Censorship. This is not an isolated fact. The geodesic and Gaussian curvatures of a $2-$ dimensional Jacobi metric can be used to study the stability of the circular orbits. The Gaussian curvature of the generalized Jacobi metric, although looking very different  from the Jacobi metric, inherits the expected properties, such as the near horizon and the asymptotic limits.
It could be interesting to extend the formalism for spacetimes with different asymptotic properties. It seems that the generalization is direct but since the global geometry of the spacetime is different it remains as an open problem.  We have seen that an intrinsic quantity such as the Gaussian curvature, the null case, is always negative at the horizon, and therefore it has to grow to reach zero at the asymptotic limit. These intrinsic geometric properties constitute a strong hint that the properties of a Riemannian metric can be used to study the dynamics of a Lorentzian geometry.  This is an important aspect of our approach, and it shows that somehow the Lorentzian properties of the original metric can be inherited  by the Riemannian metric.

Our proposal also shows clearly the relationship between the conserved quantities, they are still conserved when the Jacobi metric is projected. The new recipe will help to address different problems such as the Kepler problem in stationary spacetimes. Our formalism will help to study the geodesic motion for any asymptotically flat, stationary axisymmetric, spherically symmetric spacetimes using the tools of the Riemannian geometry.

\section{Acknowledgments}
We thank Pablo Bueno for useful discussions. The author Marcos A. Argañaraz acknowledge support from CONICET, SeCyT-UNC and FONCYT.

\bigskip

\appendix

\section{\label{app:1} Gaussian curvature of the generalized Jacobi metric}

The Gaussian curvature in two dimensions is proportional to the scalar curvature of the manifold. Here we present the Gaussian curvature of the generalized Jacobi metric \eqref{2dJac}. In order to find the extrema of this curvature we have to  differentiate with respect to the radial coordinate $r$ and solve the equation for $r$. Thus, here we present the  Gaussian curvature as a function of $f(r)$:

\begin{equation}\label{gaussianc}
K_{G}=\frac{1}{2}\frac{Q_{G}}{4 r^2 f(r)^2 (L^2 f(r)\csc^2(\theta)-r^2(\delta f(r)+E^2))^2},
\end{equation}

where

\begin{equation}
\begin{split}
Q_{G}&=\biggl[r^2 f(r)+1\biggr]\times{}\\
&\phantom{{}={}}\biggl[L^2 r^2 f(r) \csc ^2(\theta ) \biggl(2 f(r)^2 (\delta r  f'(r)-4 \delta+4 E^2)+8 \delta f(r)^3-2E^2 f(r)(r^2 f''(r)+4 r f'(r)+4)\\
&+3 E^2 r^2 f'(r)^2\biggr)\\
&\phantom{{}={}}+r^5\biggl(f'(r)(E^2 (-r)f'(r) (3 \delta f(r)+E^2)\delta f(r)^2 ( f(r)+E^2))+2 E^2 r f(r) f''(r) (\delta f(r)+E^2)\biggr)\\
 &\phantom{{}={}}  +12 L^2 r^2 f(r)^2 \csc^4(\theta)(\delta f(r)+E^2)-4 L^4 f(r)^3 \csc ^6(\theta )\biggr].
\end{split}
\end{equation}


\begin{thebibliography}{9}

\bibitem{Pin:1975}  
O.~Chong Pin, ``Curvature and mechanics", Advances in Mathematics, 15, 3, 269-311, (1975).

\bibitem{Gibbons:2015qja}
G.~W.~Gibbons,
``The Jacobi-metric for timelike geodesics in static spacetimes'',
Class.\ Quant.\ Grav.\  {\bf 33} (2016) no.2,  025004
arXiv:\arxiv{1508.06755}

\bibitem{Szydlowski:1996}
M. ~Szydlowski, Geometry of spaces with the Jacobi metric, Journal of Mathematical Physics 37, 346 (1996).


\bibitem{Chanda:2016aph}
S.~Chanda, G.~W.~Gibbons and P.~Guha,
``Jacobi-Maupertuis-Eisenhart metric and geodesic flows,''
J.\ Math.\ Phys.\  {\bf 58} (2017) no.3,  032503,
\doi{10.1063/1.4978333},
arXiv:\arxiv{1612.00375}.
 
\bibitem{Chanda:2019guf}
S.~Chanda, G.~W.~Gibbons, P.~Guha, P.~Maraner and M.~C.~Werner,
``Jacobi-Maupertuis Randers-Finsler metric for curved spaces and the gravitational magnetoelectric effect,''
J. Math. Phys. \textbf{60} (2019) no.12, 122501,
\doi{10.1063/1.5098869},
arXiv:\arxiv{1903.11805} [gr-qc].
  
  \bibitem{Duenas-Vidal:2022kcx}
\'A.~Duenas-Vidal and O.~Lasso Andino,
``The Jacobi metric approach for dynamical wormholes,''
Gen. Rel. Grav. \textbf{55} (2023) no.1, 9
\doi{10.1007/s10714-022-03060-w},
arXiv:\arxiv{2212.14147}[gr-qc].
  
   
\bibitem{Das:2016opi}
P.~Das, R.~Sk and S.~Ghosh,``Motion of charged particle in Reissner–Nordström spacetime: a Jacobi-metric approach,''
Eur.\ Phys.\ J.\ C {\bf 77} (2017) no.11,  735,
\doi{10.1140/epjc/s10052-017-5295-6},
arXiv:\arxiv{1609.04577}.

\bibitem{Arganaraz:2019fup}
M.~Arga\~naraz and O.~Lasso Andino,
``Dynamics in wormhole spacetimes: a Jacobi metric approach,''
Class. Quant. Grav. \textbf{38} (2021) no.4, 045004,
\doi{10.1088/1361-6382/abcf86},
arXiv:\arxiv{1906.11779} [gr-qc].

\bibitem{Wald:1984rg}
R.~M.~Wald,``General Relativity,'' Chicago Univ. Pr., 1984,
\doi{10.7208/chicago/9780226870373.001.0001}.

\bibitem{Chanda:2016sjg}
S.~Chanda, G.~W.~Gibbons and P.~Guha,
 ``Jacobi–Maupertuis metric and Kepler equation,''
  Int.\ J.\ Geom.\ Meth.\ Mod.\ Phys.\  {\bf 14} (2017) no.07,  1730002,
\doi{doi:10.1142/S0219887817300021},
arXiv:\arxiv{1612.07395 [math-ph]}.



\bibitem{Cardoso:2008bp}
V.~Cardoso, A.~S.~Miranda, E.~Berti, H.~Witek and V.~T.~Zanchin,
``Geodesic stability, Lyapunov exponents and quasinormal modes,''
Phys. Rev. D \textbf{79} (2009) no.6, 064016
\doi{10.1103/PhysRevD.79.064016},
arXiv:\arxiv{0812.1806}[hep-th].

\bibitem{Rahman:2018oso}
M.~Rahman, S.~Chakraborty, S.~SenGupta and A.~A.~Sen,
``Fate of Strong Cosmic Censorship Conjecture in Presence of Higher Spacetime Dimensions,''
JHEP \textbf{03} (2019), 178,
\doi{10.1007/JHEP03(2019)178}
arXiv:\arxiv{1811.08538}[gr-qc].
  


\bibitem{Cunha:2022nyw}
P.~V.~P.~Cunha, C.~A.~R.~Herdeiro and J.~P.~A.~Novo,
``Null and timelike circular orbits from equivalent 2D metrics,''
Class. Quant. Grav. \textbf{39} (2022) no.22, 225007
\doi{10.1088/1361-6382/ac987e},
arXiv:\arxiv{2207.14506}[gr-qc].



\bibitem{Tsiganov:2001}
~A.~V. Tsiganov
``The Maupertuis Principle and Canonical Transformations of the Extended Phase Space,''
J. Nonlinear Math. Phys. 8 (2001), no. 1, 157-182;
\doi{10.2991/jnmp.2001.8.1.12},
arXiv:\arxiv{nlin/0101061}[nlin.SI].


  

  
  


  




\end{thebibliography}
\end{document}